\documentclass{aa}  

\usepackage{graphicx}
\usepackage{txfonts}
\usepackage{mathrsfs}
\usepackage[colorlinks,allcolors=blue]{hyperref}
\usepackage{breakurl}
\usepackage{subfigure}

\def \src {\mbox{IGR\,J08408$-$4503}}
\begin{document} 

\title{Accretion disc by Roche lobe overflow in the supergiant fast X-ray transient \src}


 \author{L. Ducci
          \inst{1}
          \and
          P. Romano\inst{2}
          \and
          L. Ji\inst{1}
          \and
          A. Santangelo\inst{1}
          }

   \institute{Institut f\"ur Astronomie und Astrophysik, Kepler Center for Astro and Particle Physics, Eberhard Karls Universit\"at, 
              Sand 1, 72076 T\"ubingen, Germany\\
              \email{ducci@astro.uni-tuebingen.de}
              \and 
              INAF -- Osservatorio Astronomico di Brera, via Bianchi 46, 23807 Merate (LC), Italy
             }

   \date{Received ...; accepted ...}
  \abstract
   {Supergiant fast X-ray transients (SFXTs) are X-ray binary systems with a supergiant companion
    and likely a neutron star, which show a fast ($\sim 10^3$\,s) and high variability
    with a dynamic range up to $10^{5-6}$. Given their extreme properties, 
    they are considered among the most valuable laboratories to test accretion models.
    Recently, the orbital parameters of a member of this class, \src, 
    were obtained from optical observations.
    We used this information, together with X-ray observations from previous publications 
    and new results from X-ray and optical data
    collected by \emph{INTEGRAL} and presented in this work, to study the accretion mechanisms at work in \src.
    We found that the high eccentricity of the compact object orbit and the large size of the donor star imply
    Roche lobe overflow (RLO) around the periastron.
    It is also likely that a fraction of the outer layers of the photosphere of the donor star 
    are lost from the Lagrangian point $L_2$ during the periastron passages.
    On the basis of these findings, we discuss the flaring variability 
    of \src\ assuming the presence of an accretion disc.
    We point out that \src\ may not be the only SFXT with an accretion disc fueled by RLO.
    These findings open a new scenario for accretion mechanisms in SFXTs,
    since most of them have so far been based on the assumption of spherically symmetric accretion.
   }


   \maketitle
%

   \section{Introduction}
   
Supergiant fast X-ray transients (SFXTs) are a subclass of high mass X-ray binaries (HMXBs)
composed of a compact object, that is a neutron star (NS) or black hole, accreting material from an OB supergiant star
(for recent reviews, see \citealt{Romano15} and \citealt{Sidoli17}).
Classical HMXBs with a supergiant companion are X-ray sources with a typical X-ray luminosity
of $\sim 10^{35-37}$\,erg\,s$^{-1}$ and moderate variability, which have a flux dynamic range of about ten.
In contrast, SFXTs show sporadic X-ray outbursts with a duration of the order of about a day,
composed of short ($\sim 10^2-10^3$\,s) and bright ($\sim 10^{35-38}$\,erg\,s$^{-1}$) flares.
The low luminosity state is characterised by $L_{\rm low} \approx 10^{32-33}$\,erg\,s$^{-1}$, 
and therefore the flux dynamic range can reach $10^5-10^6$. These SFXTs show X-ray spectra reminiscent of those of accreting X-ray pulsars (power law
with cut-off in the range 10$-$30\,keV), indicating a likely NS 
nature for the compact object of the binary system.
Several accretion mechanisms have been proposed to explain the X-ray variability
of SFXTs. These mechanisms involve the accretion of wind clumps onto the NS 
(e.g. \citealt{intZand05,Ducci09,Chaty13}, and references therein),
gating mechanisms \citep{Grebenev07,Bozzo08}, subsonic settling accretion regimes (e.g. \citealt{Shakura14}),
and formation of transient accretion discs \citep{Ducci10}.
\src\ has the lowest duty cycle among the SFXTs,
with an activity duty cycle of 0.11\% (according to the definition of \citealt{Paizis14}
based on \emph{INTEGRAL} data)
or, equivalently, highest inactivity duty cycle of $67.2^{+4.9}_{-5.7}$\%
(according to the definition\footnote{The duty cycle of inactivity is 
   the time each source spends undetected down to a flux limit
   of 1--3$\times10^{-12}$ erg cm$^{-2}$ s$^{-1}$, the limiting flux for a 1\,ks XRT exposure.}
of \citealt[][]{Romano09b,Romano14c} based on Swift/X-Ray Telescope (XRT) data).
\src\ was discovered on 2006 May 15 with the International Gamma-Ray Astronomy Laboratory (\emph{INTEGRAL}) \citep{Goetz06}.
This system hosts a compact object orbiting around HD\,74194, a O8.5\,Ib-II(f)p
supergiant \citep{Sota14}, has an orbital period of $P=9.5436\pm 0.0002$\,d, and a high eccentricity
of $e=0.63\pm0.03$ \citep{Gamen15}.
\noindent Its distance from the Sun, derived from the parallax measurements 
in the \emph{Gaia} Data Release 2, 
is $d=2.21{+0.14\atop -0.16}$ kpc \citep{Bailer-Jones18}.
\src\ shows, similarly to other SFXTs, sporadic X-ray
outbursts composed by flares reaching X-ray luminosities of $L_{\rm x}\approx 10^{36}$\,erg\,s$^{-1}$
(e.g. \citealt{Romano09}). 
During the quiescence, its X-ray luminosity is of the order of $L_{\rm x}\approx 10^{32-33}$\,erg\,s$^{-1}$ \citep{Sidoli10,Bozzo10}.
The X-ray spectrum in the energy range 0.5-50\,keV during the flares can be described by an absorbed power law with high-energy cut-off and a black-body component (\citealt{Sidoli09}, \citealt{Romano09}, \citealt{Goetz07}, \citealt{Leyder07}).
No cyclotron resonance scattering features have been detected at hard energies.
In quiescence, the 0.3-10\,keV spectrum is well described by a soft thermal plasma model and a hard power law, 
with two different absorptions, indicating
that \src\ also accretes at low luminosity \citep{Sidoli10,Bozzo10}.
Mechanisms involving accretion of inhomogeneous stellar winds \citep{Romano09},
and gating mechanisms \citep{Goetz07,Bozzo10}
have been proposed for this source.
In light of the orbital parameters recently derived by \citet{Gamen15} from spectroscopic
optical observations, we investigated the possible accretion mechanisms responsible for the
X-ray variability exhibited by \src.
For this work we made use of \emph{INTEGRAL}, \emph{Swift}/Burst Alert Telescope (BAT), Monitor of All-sky X-ray Image (\emph{MAXI}), \emph{Suzaku},
and \emph{XMM-Newton} data previously published
\citep{Goetz07,Leyder07,Romano09,Sidoli09,Bozzo10,Sidoli10}
and  of \emph{INTEGRAL} data analysed in this work
for the first time (Sect. \ref{sect. data}). The results are discussed in Sect. \ref{sect. res disc}.

\section{Observations of \src}
\label{sect. data}

\subsection{Collection of previous X-ray observations} 

\begin{figure}[ht!]
   \centering
   \includegraphics[width=\columnwidth]{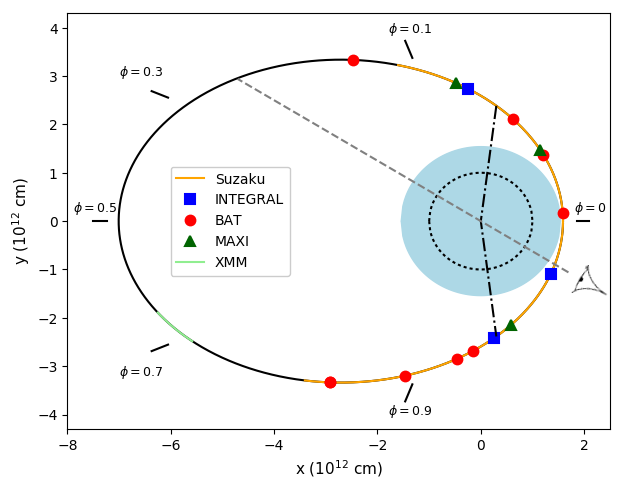}
   \caption{Positions of the bright X-ray flares ($L_{\rm x} \gtrsim 10^{36}$\,erg\,s$^{-1}$) 
     and the \emph{Suzaku} and \emph{XMM-Newton} observations
     along the orbit of \src. The blue circle shows the size of the donor star ($R_{\rm d}=22$\,R$_\odot$),
     the dotted circle shows the size of the Roche lobe at the periastron, and the dot-dashed line 
     shows the fraction of the orbit around the periastron where the system is in RLO.}
         \label{fig1}
   \end{figure}

We calculated the orbital phases and the position along the orbit of all the bright flares 
($L_{\rm x} \gtrsim 10^{36}$\,erg\,s$^{-1}$) of \src\ observed since its discovery 
by \emph{Swift}/BAT, \emph{INTEGRAL}, and \emph{MAXI}.
For \emph{INTEGRAL}, we also searched for previously unreported flares in all the archival data 
and we found one (see Sect. \ref{sect integral analysis}).
In addition, we considered the long monitoring (about three days) performed by \emph{Suzaku},
where \src\ was observed accreting at an X-ray luminosity of $10^{32}-10^{33}$\,erg\,s$^{-1}$ \citep{Sidoli10},
and an \emph{XMM-Newton} observation where \src\ was caught at a similar low luminosity level \citep{Bozzo10}.
We assumed the orbital parameters derived by \citet{Gamen15},
and we also kept their assumption about the mass of the donor star, $M_{\rm d}=33$\,M$_\odot$ 
(based on the calibration work of \citealt{Martins05}) and their derivation
for the mass of the NS, $M_{\rm ns}=1.61$\,M$_\odot$.
The times and orbital phases of these observations are shown in Table \ref{Table results1}.
The positions of these observations along the orbit are shown in Fig. \ref{fig1}.

\begin{table}
\begin{center}
\caption{Bright X-ray flares ($L_{\rm x} \gtrsim 10^{36}$\,erg\,s$^{-1}$)
detected by \emph{INTEGRAL}, \emph{Swift}/BAT, \emph{MAXI}, and long
\emph{Suzaku} and \emph{XMM-Newton} observations of \src\ at lower luminosities.}
\vspace{-0.3cm}
\label{Table results1}
\resizebox{\columnwidth}{!}{
\begin{tabular}{lccc}
\hline
\hline
\noalign{\smallskip}
Time       &  Instrument   & Orbital & Reference\\
(MJD)      &               & phase   &  \\
\noalign{\smallskip}
\hline
\noalign{\smallskip}
52821.839   & \emph{INTEGRAL}    &    0.071  & $^a$ \\
53870.770   & \emph{INTEGRAL}    &    0.980  & $^a$ \\
58489.526   & \emph{INTEGRAL}    &    0.944  & this work \\
53853.223   & \emph{Swift}/BAT &    0.141  & $^b$ \\
54012.613   & \emph{Swift}/BAT &    0.843  & $^b$  \\
54214.559   & \emph{Swift}/BAT &    0.003  & $^b$  \\
54652.887   & \emph{Swift}/BAT &    0.932  & $^b$  \\
54730.320   & \emph{Swift}/BAT &    0.046  & $^b$  \\
55071.953   & \emph{Swift}/BAT &    0.843  & $^b$  \\
55283.664   & \emph{Swift}/BAT &    0.026  & $^b$  \\
55798.035   & \emph{Swift}/BAT &    0.923  & $^b$  \\
56475.341   & \emph{Swift}/BAT &    0.893  & $^c$ \\
55914.029   & \emph{MAXI}      &    0.077  & $^d$ \\
57640.955   & \emph{MAXI}      &    0.028  & $^d$ \\
58451.443   & \emph{MAXI}      &    0.953  & $^d$ \\
55176.746-55179.500  & \emph{Suzaku} & $-0.177\leq \phi \leq+0.112$ & $^e$ \\
54249.391-54249.918  &\emph{XMM-Newton} & $0.653\leq \phi \leq 0.708$ & $^f$ \\
\hline
\end{tabular}
}
\end{center}
Notes. $^a$: \citet{Goetz07}; $^b$: \citet{Romano14}; $^c$: \citet{Romano13};
$^d$: List of the detection of transient sources reported in \url{http://maxi.riken.jp/alert/novae/index.html} ;
$^e$: \citet{Sidoli10}; $^f$: \citet{Bozzo10}.
\end{table}

\subsection{Analysis of INTEGRAL data}
\label{sect integral analysis}

\begin{figure}[ht!]
   \centering
   \includegraphics[width=\columnwidth]{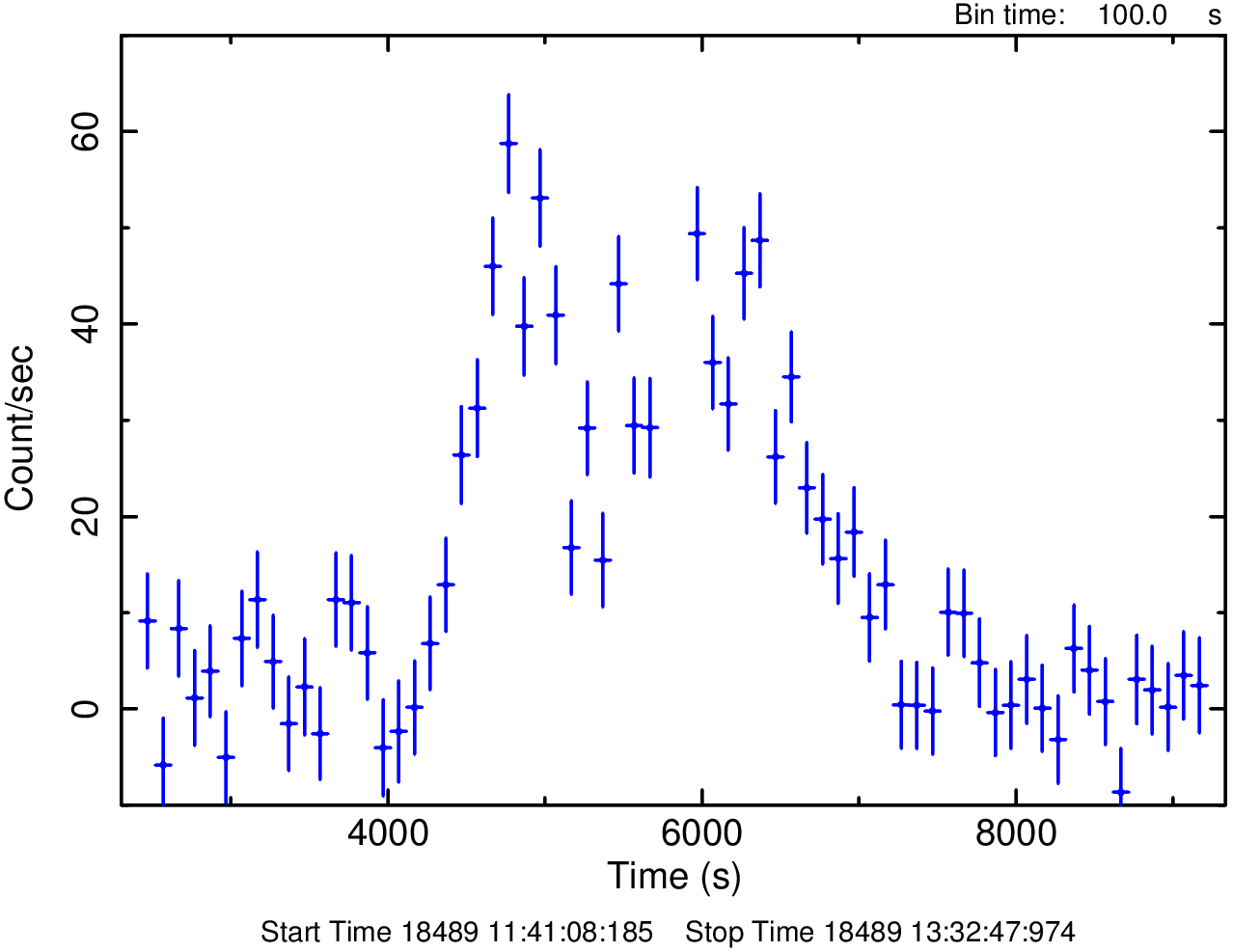}\\
   \includegraphics[width=\columnwidth]{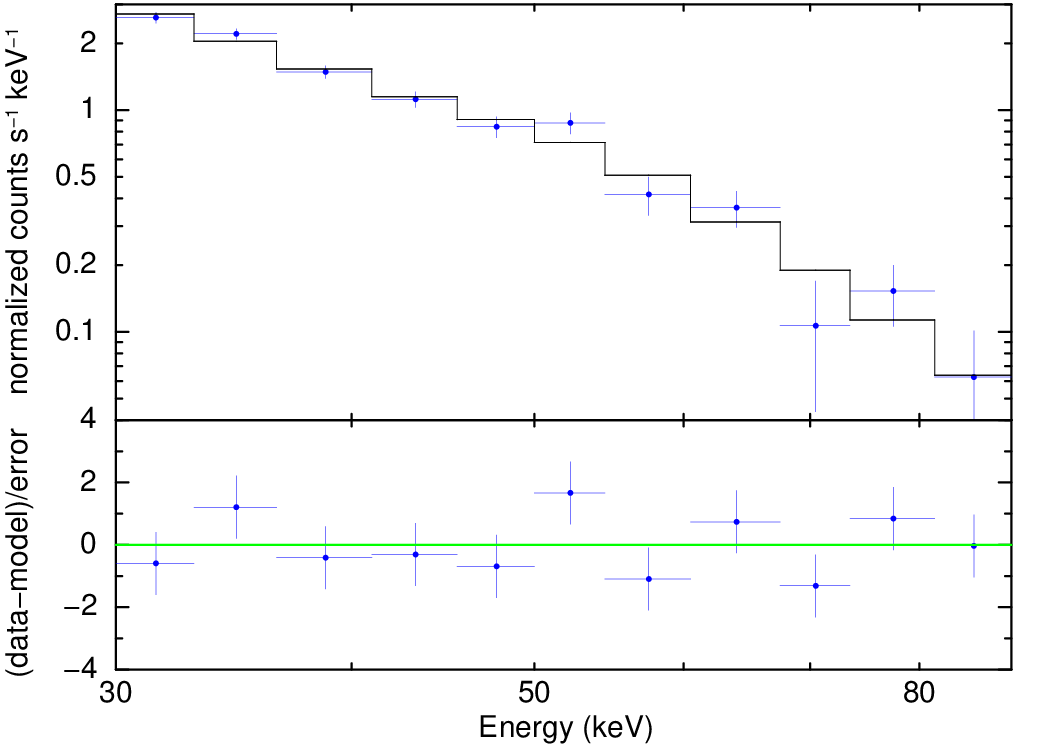} 
   \caption{\emph{Top panel:} 30$-$50\,keV light curve of the 2019 January flare of \src.
            \emph{Bottom panel:} IBIS/ISGRI spectrum of the flare of \src.}
         \label{fig2}
   \end{figure}

We analysed all the available archival data collected by the
INTEGRAL Soft Gamma-Ray Imager (ISGRI; \citealt{Lebrun03}) detector of the
Imager on Board INTEGRAL Satellite (IBIS; \citealt{Ubertini03})
on board INTEGRAL \citep{Winkler03},
from 2003 January 11 to 2019 February 17 (corresponding to 3222 pointings).
The IBIS instrument operates in the $\sim 20$\,keV$-10$\,MeV energy range.
We performed the analysis using the Off-line Science Analysis (OSA)
software provided by the ISDC Data Center for Astrophysics\footnote{OSA v.\,11 for data collected after
\emph{INTEGRAL} revolution 1626: \url{https://www.isdc.unige.ch/integral/download/osa/doc/11.0/osa_um_ibis/man.html},
and OSA v.\,10.2 otherwise: \url{https://www.isdc.unige.ch/integral/download/osa/doc/10.2/osa_um_ibis/man.html}}.
We only considered observations with \src\ within $12^\circ$ from the centre of the
IBIS/ISGRI field of view (FOV).
In addition to the previously known flares, \src\ showed a bright flare in 2019 January 6 at 12:11:31.200 UTC 
(see Fig. \ref{fig2}, top panel).
During the flare ($T_{\rm start}=58489.508$\,MJD; $T_{\rm stop}=58489.541$\,MJD) the spectrum can be described
by a a power law with high energy cut-off, with photon index of $2.5 \pm 0.2$, $E_{\rm c}=54.0{+8.9 \atop-8.8}$\,keV,
$E_{\rm f}=34{+31 \atop -16}$\,keV ($\chi^2_{\rm red}=1.36$, 7 d.o.f.) Uncertainties are given at 68\% confidence level. We added systematic uncertainties of 3\% to the IBIS/ISGRI spectrum.
The peak flux is $\sim 2.7\times 10^{-9}$\,erg\,cm$^{-2}$\,s$^{-1}$ (30$-$80\,keV),
corresponding to a luminosity of $L_{\rm x}\approx 1.6\times 10^{36}$\,erg\,s$^{-1}$ (Fig. \ref{fig2}, bottom panel).
We searched for periodicities using IBIS/ISGRI data during the flare 
(i.e. we defined a good time interval with $T_{\rm start}-T_{\rm stop}$
defined above) by selecting events with pixel fraction illuminated by \src\ equal to one
from the detector shadowgrams obtained before the reconstruction (deconvolution)
of the sky image.
In addition, we also performed this search using binned light curves (bin$=0.1$\,s)
obtained with the OSA binning tool \emph{ii\_light}.
In both cases the Solar System barycentre correction was applied.
The search for periodic signals was performed in the range 0.01$-$1000\,s with the event files
and 0.2$-$1000\,s with the binned light curve.
We used the $Z^2$ and the Lomb-Scargle periodogram techniques \citep{Scargle82,Lomb76,Buccheri83}.
No statistically significant periodicity was detected.
During this flare, \src\ was outside of the JEM-X FOV.
We estimated a 2$\sigma$ upper limit to the pulsed fraction of 55\% and 50\% for periods in the ranges 0.2$-$10\,s
and 10$-$1000\,s, respectively.

\begin{figure}[ht!]
   \centering
   \includegraphics[width=\columnwidth]{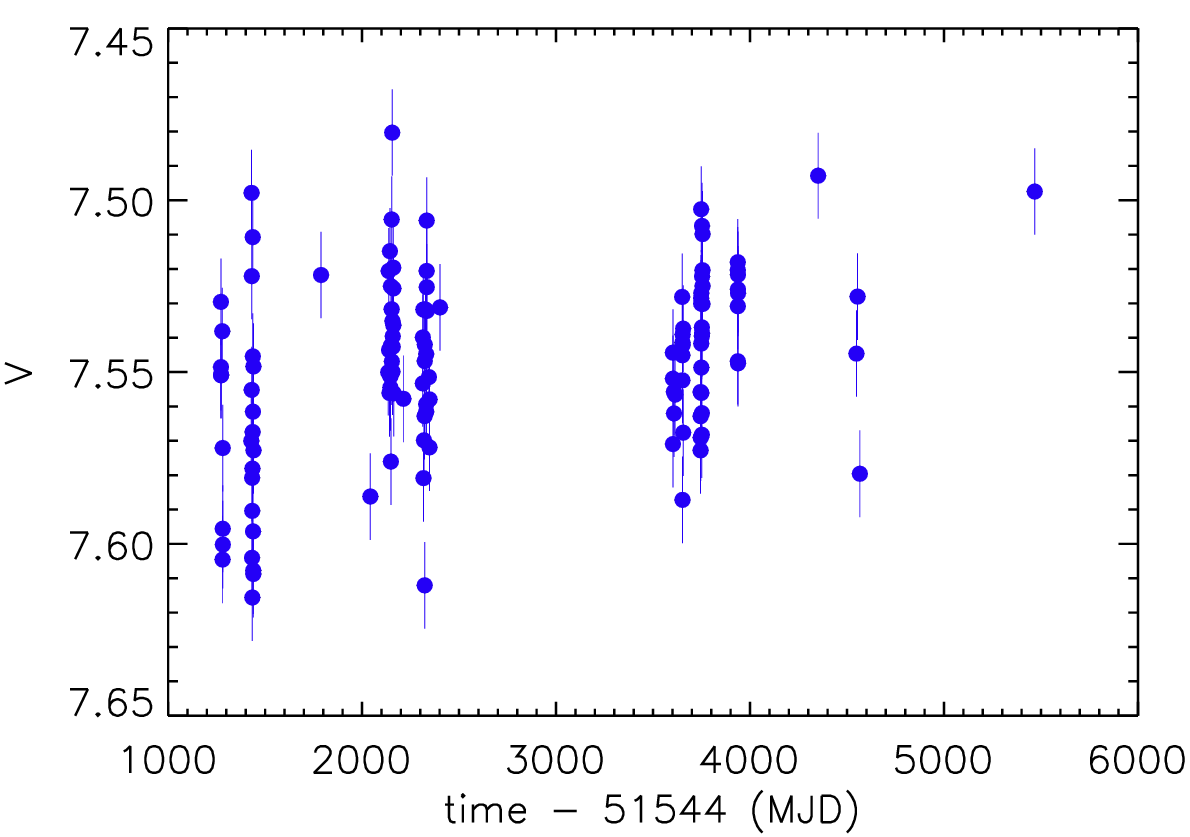}\\
   \includegraphics[width=\columnwidth]{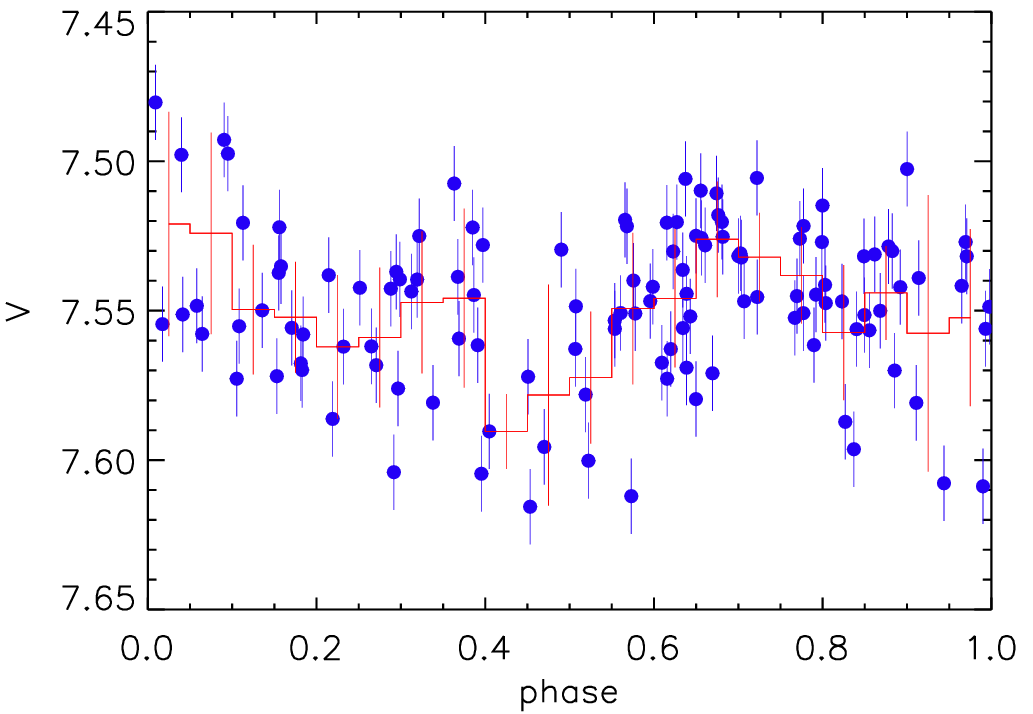}
   \caption{\emph{Top panel:} $V$ band light curve of \src, observed by the OMC.
     \emph{Bottom panel:} The same light curve, folded with the orbital period found by \citet{Gamen15}. Phase zero corresponds to the periastron passage. Blue points have a time duration of 10\,s.
   Red line shows the binned light curve. The error bars of the red line are the standard deviation obtained from the magnitude values of the blue points and show their dispersion.}
         \label{fig3}
   \end{figure}

We performed an analysis of all the available 
photometric data of \src\ collected by the Optical Monitoring Camera 
(OMC; \citealt{Mas-Hesse03}) on board \emph{INTEGRAL}.
This camera has a much smaller FOV compared to ISGRI (4.979$^\circ\times$4.979$^\circ$) and it is equipped 
with a Johnson $V$ filter.
We reduced the data following the prescriptions of the OMC Analysis User Manual\footnote{\url{https://www.isdc.unige.ch/integral/download/osa/doc/11.0/osa_um_omc.pdf}}.
Since \src\ is bright ($V\approx 7.5$), 
we considered only the observations with integration time of 10\,s
because they allow a good measurement of the magnitude without pixel saturation.
We thus selected 124 ``good'' observations of \src.
Figure \ref{fig3} shows the light curve as a function of time (top panel) and folded with the orbital period
of the source (bottom panel), where phase zero corresponds to the periastron passage.
The OMC light curve presented here is, to the best of our knowledge, the most detailed
and long photometric monitoring of this source.
It shows irregular variability with amplitudes up to $\sim 0.14$.

   \section{Results and discussion} \label{sect. res disc}

The bright X-ray flares of \src\ previously observed and the last flare detected by \emph{INTEGRAL}
and reported in this work are clustered around the periastron, within the orbital interval $\sim 0.84-0.14$.
The long \emph{Suzaku} observation carried out during the same orbital phases of the flares
shows a variable low luminosity state of $\sim 10^{32-33}$\,erg\,s$^{-1}$ \citep{Sidoli10},
which implies a luminosity dynamic range of $\Gamma=10^3-10^4$.
Based on the calibration of O star parameters provided by \citet{Martins05},
we found that the average radius of the donor star of \src, $R_{\rm d}\approx 22$\,R$_\odot$,
is larger than the size of its Roche lobe at the periastron, $R_{\rm l,d}\approx 14.4$\,R$_\odot$,
calculated with the formula provided by \citet[][equation 2]{Eggleton83} assuming the stellar and orbital
parameters previously adopted in this work (see Fig. \ref{fig1}, 
where the dotted circle shows the $R_{\rm l,d}$ at periastron).
The dot-dashed line of Fig. \ref{fig1} shows the fraction of the orbit where
the system is in Roche lobe overflow (RLO; i.e. when $R_{\rm d}\gtrsim R_{\rm l,d}$).
We obtained similar results using the calibration of O star parameters
provided by \citet{Vacca96}, although in this case the radius of the donor star
results to be larger than the periastron distance.
The strong tidal interaction that would be caused by the periodic passage of the NS
inside the photosphere of the companion star would likely produce a displacement
of a significant fraction of the photosphere itself. Consequently, this would lead
to a periodic optical luminosity and column density variability.
In addition, the high mass accretion rate at $\phi=0$, where the NS would
be inside the companion star, would cause a higher rate of X-ray flares
at $\phi \approx 0$. Since none of these properties has been observed in \src,
this suggests that the radius of the donor star is likely to be smaller than the periastron distance.
With the information about the visual magnitude measured by the OMC
  ($m_{\rm v} = 7.55 \pm 0.05$), the  distance ($d=2.21{+0.14\atop-0.16}$\,kpc),
the effective temperature ($T_{\rm eff}=32274\pm1800$\,K, \citealt{Martins05})\footnote{The error on $T_{\rm eff}$
is a rough estimate obtained by the values in tables 3 and 6 of \citet{Martins05}.},
the absorption ($N_{\rm H}=5.5\pm0.4\times 10^{21}$\,cm$^{-2}$)\footnote{Obtained from the \emph{XMM-Newton}
observations of \src\ during the low luminosity state and reported in \citet{Sidoli13}.
We only considered the absorption affecting the MEKAL (Mewe-Kaastra-Liedahl) component
to exclude the absorption produced around the accreting compact object.
The column density is obtained from the weighted mean of the values reported in table 3 of \citet{Sidoli13},
with uncertainties corrected to 1$\sigma$ confidence level.
We converted $N_{\rm H}$ to $A_{\rm V}$ using the formula presented in \citet{Guever09}
and taking into account the uncertainties reported in their paper.},
and assuming that the radiation from the donor star is a black body,
we found a radius of $R_{\rm d}=23.4{+7.0\atop-5.4}$\,R$_\odot$.
Although it is not well constrained, we note that it is consistent with the values of $R_{\rm d}$
mentioned above for this type of stars, and thus consistent with the RLO hypothesis.
Using the $B$ and $R$ magnitudes from the United States Naval Observatory B1 (USNO\,B1) catalogue \citep{Monet03}
  and the extinction law of \citet{Cardelli89}, we obtained $R_{\rm d}=23.3{+10.1\atop -8.3}$\,R$_\odot$.
  The larger uncertainty of this estimate is due to the worst photometric accuracy of the USNO\,B1 catalogue with respect to the
OMC measurements.

\citet{Regos05} studied the mass transfer in eccentric binaries 
in which RLO is expected to occur.
They found that for high eccentricities ($e\geq 0.6$), a high mass transfer at periastron is possible,
and matter leaves the donor star through the Lagrangian points $L_1$ and $L_2$.
A fraction of the matter leaving $L_2$ falls back to the star, while the other part 
does not remain bound to the star. 
Similarly, part of the stream passing through $L_1$ leaves the system,
while most of it forms an accretion disc.
Given the orbital parameters,
this mechanism might be at work in \src.
The stream of matter lost by the donor star through $L_2$ could make the environment 
around the binary system not ``cleaned'': slow and denser regions of gas could be crossed
by the NS along its orbit, feeding further the NS.
In this scenario, the aperiodic optical variability shown by \src\ in the OMC light curve
(Fig. \ref{fig3}) might be caused by sporadic absorption of dense clumps of gas lost from $L_1$ and $L_2$
crossing the line of sight, or by tidal interactions between the two stars, which 
disrupts the outer layers of the photosphere of HD\,74194.
As previously mentioned, another possible peculiarity of \src\ is the short
distance between the NS and the surface of the donor star at periastron,
that is just $\approx 6\times 10^{10}$\,cm, assuming $R_{\rm d}=22$\,R$_\odot$
(the periastron distance is $\sim 22.9$\,R$_\odot$).
This would imply that there is likely not enough space for an accretion disc which, once formed
during the orbital phases preceding $\phi \approx 0$, might
be disrupted because of the tidal interactions with the donor star.
A slightly smaller radius for the supergiant would increase the chances of survival of the disc;
for example, if $R_{\rm d}\approx 19$\,R$_\odot$, i.e.  the radius of HD\,75211, a O8.5\,II(f) star (\citealt{Markova18}).

If the NS of \src\ is fed by an accretion disc, the high X-ray variability could be caused by
transitions from inhibition of accretion (during low luminosity states) to the accretion regime (during flares).
This would be possible for small variations of the mass rate gravitationally captured by the NS ($\dot{M}_{\rm c}$)
if the corotation radius
\begin{equation} \label{r_co}
r_{\rm co}=[GM_{\rm ns}P_{\rm spin}^2/(4\pi^2)]^{1/3} \mbox{ ,}
\end{equation}
defined as the distance from the NS where material in a Keplerian orbit corotates with the NS,
is comparable in size to that of the inner radius of the disc; this is defined as the distance from the NS where
the magnetic pressure and pressure exerted by the matter in the disc are balanced: 
\begin{equation} \label{r_in}
  r_{\rm in} \approx \xi [\mu^4/(GM_{\rm ns}\dot{M}_{\rm c}^2)]^{1/7} \mbox{ .}
\end{equation}
In these equations, $M_{\rm ns}$ is the mass of the NS and $P_{\rm spin}$ its spin period,  
$\xi\approx 0.4$ in case of accretion from a Keplerian disc or 1
in case of spherical accretion (see \citealt{Revnivtsev15}
and references therein), $\mu$ is the magnetic dipole moment, $B=2\mu/R_{\rm ns}^3$ is the magnetic field
strength at the poles, and $R_{\rm ns}$ is the radius of the NS.

If $r_{\rm in} > r_{\rm co}$, the centrifugal barrier sets in.
During this stage the accretion is expected to be negligible.
Once $r_{\rm in} \approx r_{\rm co}$, the centrifugal barrier is overcome and accretion sets in.
In this framework, we can consider the following very simplified calculations.
We assume that during the accretion state the X-ray luminosity is given by
$L_{\rm acc}\approx GM_{\rm ns}\dot{M}_{\rm c}/R_{\rm ns}$. When the magnetic barrier activates
($r_{\rm in}=r_{\rm co}$) for small variations of $\dot{M}_{\rm c}$, which for simplicity can be considered constant
in our calculations, this value reduces to
$L_{\rm gate}\approx GM_{\rm ns}\dot{M}_{\rm c}/r_{\rm co}$.
Therefore, $r_{\rm co}$ can be expressed as a function of $\Gamma=L_{\rm acc}/L_{\rm gate}$,
\begin{equation} \label{r_co vs gamma}
r_{\rm co} = R_{\rm ns}\Gamma \mbox{ .}
\end{equation}
Assuming again $r_{\rm in}=r_{\rm co}$, from Eq. \ref{r_co vs gamma} and Eq. \ref{r_in},
we obtain the following relation between $\mu$, $L_{\rm acc}$, and $\Gamma$:
\begin{equation} \label{mu}
\mu = \xi^{-7/4} R_{\rm ns}^{9/4} \Gamma^{7/4} G^{-1/4} M_{\rm ns}^{-1/4} L_{\rm acc}^{1/2} \mbox{ .}
\end{equation}
Assuming $\Gamma=10^3-10^4$, $L_{\rm acc}=10^{36}$\,erg\,s$^{-1}$, $R_{\rm ns}=12$\,km, $M_{\rm ns}=1.61$\,M$_\odot$, and $\xi=0.4$, 
we find $\mu= 1.1\times 10^{31} - 6.2\times 10^{32}$\,G\,cm$^3$ 
and thence $B= 1.3\times 10^{13} - 7.2\times 10^{14}$\,G.
From Eq. \ref{r_co} we then obtain $P_{\rm spin}=18 - 570$\,s,
while the corotation radius is $r_{\rm co}=1.2 \times 10^9 - 1.2\times 10^{10}$\,cm.
These calculations indicate a magnetar nature for the compact object of \src\
for a wide range of values of $\Gamma$.
The lack of clear signatures for the presence of an accretion disc
  in the X-ray spectrum of \src\ in the spectral analyses presented in previous works
  is in agreement with the high magnetic field and low spin period inferred from the calculations
  presented above. If $r_{\rm in}=r_{\rm co}$, from the equation of the spectrum
  emitted by an optically thick accretion disc \citep[e.g.][equation 5.43]{FKR},
  \begin{equation} \nonumber
  T(r) = \left [ \frac{3 G M \dot{M}_{\rm c}}{8 \pi r^3 \sigma} \left ( 1 - \sqrt{\frac{R_{\rm ns}}{r}} \right ) \right ]^{1/4} \mbox{ ,}
  \end{equation}
  where $\sigma$ is the Stefan-Boltzmann constant, we obtain $T(r_{\rm in})\sim 1$\,eV,
which means that an accretion disc would be too cold to be observed in X-ray.

  A possible problem for the accretion disc scenario is that
  the bright flares ($L_{\rm x} \gtrsim 10^{36}$\,erg\,s$^{-1}$) are clustered
  in an interval of the orbit around the periastron ($\Delta \phi \approx 0.3$).
  We expect a higher mass accretion rate in this interval of the orbit 
  because of RLO and because the stellar wind is denser and slower closer to the donor star. However
  if the accretion is mediated by a disc, the NS would feel effects
  of the mass accretion rate variability with a time delay, which is 
  of the order of the viscous timescale ($\sim$ days; for \src)
  and not necessarily in phase with the orbital period.
  One possibility previously mentioned is that the size of the donor star is so large that the accretion disc
  is disrupted by tidal interactions at $\phi \approx 0$.
  In this case, the accretion is spherically symmetric along the orbit---no delay effects are expected---and the gating mechanism model of \citet{Bozzo08}
  can be applied. In this framework, the denser streams of gas expelled by the donor star
  from $L_2$ and $L_1$, and sporadically accreted by the NS during the periastron passages, could trigger the transitions
  from low to higher luminosity accretion regimes described in \citet{Bozzo08}.
  Instead, if the size of the donor star is such that the accretion disc survives at $\phi \approx 0$,
  a two-stream accretion mechanism similar to that proposed by \citet{Lipunov80,Lipunov82}
  could be in operation.
  During RLO, an accretion disc forms. In addition, the NS accretes from the
  strong stellar wind produced by the companion star all along the orbit.
  \citet{Lipunov82} pointed out that during the propeller state, which could be the low luminosity state of \src,
  the spherical stream collides with the magnetosphere and slowly settles down on the NS equatorial
  plane to form or feed an already existing accretion disc.
  In this case, the clustering of the bright flares around the periastron
  might be caused by the sudden overcoming of the centrifugal barrier ($r_{\rm in} \lesssim r_{\rm co}$)
  triggered by instabilities produced by the enhanced accretion
  produced by the spherical stream around the periastron.
  The interactions between the two accretion streams are still poorly studied.
  \citet{Mustsevoi91} pointed out that they could lead to instabilities in the accretion disc.

\src\ could not be an isolated case of accretion through RLO among SFXTs.
To the best of our knowledge there are at least two other cases: IGR\,J16418$-$4532 and IGR\,J16479$-$4514.
Their orbital, stellar, and Roche lobe properties are reported in Table \ref{Table sfxts}.
While for reasonable values of the radius, the donor star of IGR\,J16479$-$4514
could still underfill its Roche lobe (as pointed out by \citealt{Sidoli13}),
for IGR\,J16418$-$4532 the radius of the donor star is clearly too large
compared to its Roche lobe (see \citealt{Sidoli12} and \citealt{Drave13} for further discussions).
In particular, we note that with the parameters reported in Table \ref{Table sfxts},
the radius of the donor star of IGR\,J16418$-$4532 would be slightly larger than the orbital size, which is obviously not possible.
According to \citet{Crowther06}, the radii of B\,0.5Ia stars can be significantly different between different stars.
Although on average they are $\sim 33-34$\,R$_\odot$, for example another star 
of similar spectral type, $\kappa$\,Ori, has $R_*\approx 22.2$\,R$_\odot$,
which would still imply $R_*\gtrsim R_{\rm l,d}$ for IGR\,J16418$-$4532.

The formation of accretion discs through RLO could therefore be
a property common to a group of SFXTs.
This hypothesis, however, requires further investigation.
Therefore, in-depth optical observations and X-ray studies
aimed at determining more precisely the masses and radii of the SFXTs reported in Table \ref{Table sfxts}
and the presence of an accretion disc are of fundamental importance.

\begin{table}
\begin{center}
\caption{List of SFXTs which might experience RLO in a fraction or along the entire NS orbit. The quantity
$R_{\rm l,d}$ is the size of the Roche lobe of the donor star during the periastron passage.}
\vspace{-0.3cm}
\label{Table sfxts}
\resizebox{\columnwidth}{!}{
\begin{tabular}{lcccccc}
\hline
\hline
\noalign{\smallskip}
Source name        &    Sp. Type     &    Mass    &    Radius    &   $P_{\rm orb}$   &    $R_{\rm l,d}$  \\
                   &                 & (M$_\odot$) &  (R$_\odot$)  &      (d)        &     R$_\odot$    \\
\noalign{\smallskip}
\hline
\noalign{\smallskip}
IGR\,J08408$-$4503 & O8.5\,Ib-II(f)p &   33$^1$   &     22$^1$   &   9.4536$^2$    &     14.38      \\
\noalign{\smallskip}
IGR\,J16418$-$4532 & BN\,0.5Ia$^3$   &   33$^4$   &    33.8$^4$  &   3.7389$^5$    &     21.14       \\  
\noalign{\smallskip}
IGR\,J16479$-$4514 &   O9.5\,Iab$^6$    & 27.8$^1$  &    22$^1$    &  3.3193$^{7,8}$  &    18.12       \\
\noalign{\smallskip}
\hline
\end{tabular}
}
\end{center}
Notes. $^1$: \citet{Martins05}; $^2$: \citet{Gamen15}; $^3$: \citet{Coleiro13}; $^4$: \citet{Searle08}; $^5$: \citet{Levine11};
$^6$: \citet{Nespoli08}; $^7$: \citet{Jain09}; $^8$ \citet{Romano09}.
We assumed $e=0.63$ for \src\ and $e=0$ for the other SFXTs because of the lack
of an orbital modulation of the X-ray flux \citep{Sidoli12,Drave13}.
\end{table}

\begin{acknowledgements}
We thank the anonymous referee for constructive comments that
helped to improve the paper.
This paper is based on data from observations with INTEGRAL satellite,
an ESA project with instruments and science data centre funded by ESA
member states (especially the PI countries: Denmark, France, Germany, Italy,
Spain, and Switzerland), Czech Republic and Poland, and with the participation of Russia and the USA.      
This work is supported by the Bundesministerium f\"ur
Wirtschaft und Technologie through the Deutsches Zentrum
f\"ur Luft und Raumfahrt (grant FKZ 50 OG 1602).
P.\,R. acknowledges contract ASI-INAF I/004/11/0.
We acknowledge financial contribution from the agreement ASI-INAF n.2017-14-H.O.
\end{acknowledgements}

\bibliographystyle{aa} 
\bibliography{ld08408}

\end{document}